\documentclass[%
 reprint,
superscriptaddress,
 amsmath,amssymb,
 aps,
]{revtex4-2}

\usepackage{graphicx}% Include figure files
\usepackage{dcolumn}% Align table columns on decimal point
\usepackage{bm}% bold math
\begin{document}

\preprint{APS/123-QED}

\title{A nonrecursive method for computing the off-diagonal small-time heat kernel expansion}% Force line breaks with \\

\author{V.~O.~Guba}
\email{V.O.Guba@inp.nsk.su}
\affiliation{
Budker Institute of Nuclear Physics of Siberian Branch Russian Academy of Sciences, Novosibirsk, 630090 Russia
}
\affiliation{
Novosibirsk State University, Novosibirsk 630090, Russia
}
\author{A.~V.~Reznichenko}
 \email{A.V.Reznichenko@inp.nsk.su}
 \affiliation{
Budker Institute of Nuclear Physics of Siberian Branch Russian Academy of Sciences, Novosibirsk, 630090 Russia
}
\affiliation{
Novosibirsk State University, Novosibirsk 630090, Russia
}

\date{\today}% It is always \today, today,

\begin{abstract}
We present a nonrecursive method to compute the small-time heat kernel asymptotic expansion. The Klein-Gordon operator in flat space-time coupled to electromagnetic fields is considered. We obtain closed-form expressions for the expansion coefficients up to the second order in time. Our approach can be regarded as an extension of the method described in [R.I. Nepomechie, Phys. Rev. D 31, 3291 (1985)], but unlike that work, our approach is valid for arbitrary space-time dependence of the heat kernel, i.e. for off-diagonal values. We verify that our results correctly reproduce small-time asymptotics of the heat kernel for a plane wave field and for constant electromagnetic fields. 
\end{abstract}

\maketitle

%\tableofcontents

\section{\label{sec:level1}Introduction}

The heat kernel plays an important role in quantum field theory \cite{avramidi2002heat,fradkin2021quantum} and in spectral theory of linear operators \cite{davies1989heat}. For a self-adjoint operator $H$ it can be defined as the matrix element of the operator $e^{-iH\tau}$ between the eigenstates of the coordinate operator:
\begin{align}
    h(\tau,x,y) = \langle x|e^{-iH\tau}|y\rangle,
\end{align} 
where $\hat{x}_{\mu}|x\rangle = x_{\mu}|x\rangle$, $\langle x|y\rangle = \delta(x-y)$. We will work in $d$-dimensional Minkowski spacetime. In this work we consider the operator $H$ which has the following coordinate space representation $H_{x}$:
\begin{align}\label{1Hoperator}
    &H_{x} = D_{\mu}D^{\mu}+V(x),\\ \nonumber
    &D_{\mu} = \frac{\partial}{\partial x^{\mu}}+ i eA_{\mu}(x).
\end{align}
The functions (potentials) $V(x)$ and $A_{\mu}(x)$ are real. The heat kernel $h(\tau,x,y)$ is the solution to the following differential equation:
\begin{align}\label{1hequation}
    i\partial_{\tau}h(\tau,x,y) = H_{x}h(\tau,x,y),
\end{align}
with the initial condition $h(0,x,y) = \delta(x-y)$.

Operators of the type \eqref{1Hoperator} often appear in quantum field theory. As a consequence, the usage of heat kernel methods is widespread. Using the Fock-Schwinger proper-time representation it is possible, for instance, to express the propagator for the theory of a complex scalar field $\phi(x)$ with the Lagrangian density given by:
\begin{align}
    \mathcal{L} = (D_{\mu}\phi(x))^{*}(D^{\mu}\phi(x)) - V(x)|\phi(x)|^{2},
\end{align}
via the representation:
\begin{align}
    \langle x|H^{-1}|y\rangle = i\int_{0}^{\infty}d\tau \langle x |e^{-iH\tau}|y\rangle = i\int_{0}^{\infty}d\tau \,h(\tau,x,y).
\end{align}

The heat kernel is also useful for the calculation of the zeta function of the operator $H$ \cite{fradkin2021quantum}:
\begin{align}\label{1zeta}
    \zeta_{H}(s)=\frac{1}{\Gamma(s)}\int_{0}^{\infty}d\tau \, \tau^{s-1}\int d^{d}x\lim_{y\to x}h(\tau,x,y).
\end{align}
Equation \eqref{1zeta} only gives $\zeta_{H}(s)$ for those $s$ for which the integrals converge. Otherwise, the function $\zeta_{H}(s)$ is obtained by the analytic continuation. Another useful object in QFT related with the heat kernel is the regularized functional determinant, which is defined by (see \cite{schwarz2013quantum}):
\begin{align}
    \mathrm{Det}\, H = e^{-\zeta_{H}^{\prime}(0)}.
\end{align}
Functional determinants appear in calculations of Gaussian path integrals. An example of a quantity computed in terms of functional determinants is the effective action, which in one-loop order is expressed as a logarithm of a functional determinant (see \cite{avramidi2002heat, fradkin2021quantum}).

A closed analytical form of $h(\tau,x,y)$ can be obtained only for some specific choices of $V(x)$ and $A_{\mu}(x)$. For general potentials one has to use approximate expressions for the heat kernel. A widely used approximation is the asymptotic expansion of $h(\tau,x,y)$ as $\tau\to0$, which has the form:
\begin{align}\label{1hexpansionform}
    h(\tau,x,y) = \frac{ie^{-iz^{2}/(4\tau)}}{(4\pi i\tau)^{d/2}}\sum_{n=0}^{\infty}a_{n}(x,y)\tau^{n},
\end{align}
where $z_{\mu} \equiv x_{\mu}-y_{\mu}$. The diagonal values of the coefficients $a_{n}(x,y)$, i.e. when $x=y$, are important for one-loop renormalization of field theories, as shown, for instance, in \cite{dewitt1965dynamical}.

At the moment there are many methods to obtain small-time expansion of the heat kernel, and some of the methods are reviewed in the work \cite{vassilevich2003heat}. Most methods are concerned with the calculation of the diagonal coefficients $a_{n}(x,x)$. A classic method well-suited to the calculation of diagonal values $a_{n}(x,x)$ is described in \cite{dewitt1965dynamical}. This method relies on Eq.~\eqref{1hequation} and uses Eq.~\eqref{1hexpansionform} as an ansatz to obtain recursive equations for the coefficients $a_{n}(x,y)$. 

The calculation of the non-diagonal values $a_{n}(x,y)$ is more complicated, and often $a_{n}(x,y)$ are obtained only as a series in $z_{\mu}$. A method for obtaining $a_{n}(x,y)$ without expanding in $z_{\mu}$ is given, for instance, in the work \cite{bolte2013heat}. The authors of \cite{bolte2013heat} derive a recursive relation which expresses $a_{n}(x,y)$ in terms of the action of a linear integral operator on $a_{n-1}(x,y)$.

There are also non-recursive methods for obtaining $a_{n}(x,y)$. The author of \cite{nepomechie1985calculating} describes a non-recursive calculation of $a_{n}(x,x)$ deriving and then using representation of $h(\tau,x,y)$ of the following form:
\begin{align}\label{1hdiagrep}
    &h(\tau,x,y) = \frac{e^{-iz^{2}/(4\tau)}}{(4\pi  \tau)^{d/2}}\int \frac{d^{d}q}{\pi^{d/2}}e^{iq^{2}}\nonumber\\ 
    &\times\exp \left\{-i\tau H_{x}+2\sqrt{\tau}q\cdot D\right\}1,
\end{align}
where $q\cdot D \equiv q^{\mu}D_{\mu}$, and the $1$ written to the right of the operator exponential emphasizes that operators $H_{x}$ and $q\cdot D$ act on the function $\psi(x) = 1$. The usage of Eq.~\eqref{1hdiagrep} for obtaining $a_{n}(x,x)$ is straightforward. Expanding the exponential from Eq.~\eqref{1hdiagrep} and collecting the terms proportional to the desired power of $\tau$, one then performs integration over $q_{\mu}$ and computes the action of differential operators on $\psi(x)=1$. 

In this work we present a new non-recursive method for calculation of non-diagonal values of $a_{n}(x,y)$. It is essentially an extension of the method described in \cite{nepomechie1985calculating}. In Section 2 we derive a representation of $h(\tau,x,y)$ similar to Eq.~\eqref{1hdiagrep}, but we do not restrict ourselves to diagonal part of the heat kernel. Then we demonstrate use of our representation by computing the coefficients $a_{0}(x,y)$, $a_{1}(x,y)$ and $a_{2}(x,y)$. In Section 3 we  consider specific cases of plane wave and constant electromagnetic fields to verify that our results reproduce the well-known expressions. Appendix A gives details concerning the use of the representation of the heat kernel derived in Section 2.

\section{Small time asymptotics of the heat kernel}

In this section we will obtain the perturbative expansion of $h(\tau,x,y)$ in powers of the proper time $\tau$. Our method is based on and extends the method described in the work \cite{nepomechie1985calculating}. Specifically, it is possible to calculate the expansion of the non-diagonal values of $h(\tau,x,y)$ with our method, whereas the work \cite{nepomechie1985calculating} describes the calculation of the diagonal values only.

To begin with, we use the integral representation of the delta function:
\begin{align}
    &h(\tau,x,y) = \langle x|e^{-iH\tau}|y\rangle = e^{-iH_{x}\tau}\delta(x-y)=\nonumber\\
    &=\int \frac{d^{d}k}{(2\pi)^{d}}e^{-iky}e^{-iH_{x}\tau}e^{ikx},
\end{align}
where the subscript $x$ of the operator $H_{x}$ emphasizes that this operator acts on the variables $x_{\mu}$. In the expression $e^{-iH_{x}\tau}\delta(x-y)$ the operator $e^{-iH_{x}\tau}$ acts only on the delta function, and in the next line it acts only on the function $e^{ikx}$. We proceed in the same way as \cite{nepomechie1985calculating} using the following operator identity:
\begin{align}
    e^{-ikx}D_{\mu}e^{ikx} =D_{\mu} +i k_{\mu},
\end{align}
which implies:
\begin{align}\label{2Hidentity}
    e^{-ikx}H_{x}e^{ikx} = H_{x} + 2ik^{\mu}D_{\mu}-k^{2}.
\end{align}
Using \eqref{2Hidentity} we rewrite $h(\tau,x,y)$ in the following form:
\begin{align}\label{2hkintegral}
    h(\tau,x,y) = \int \frac{d^{d}k}{(2\pi)^{d}}e^{i\tau k^{2}+ikz}e^{-iH_{x}\tau+2\tau k\cdot D}1,
\end{align}
where $z_{\mu}\equiv x_{\mu}-y_{\mu}$, and we explicitly wrote $1$ to emphasize that the exponential of the operator in Eq.~\eqref{2hkintegral} acts on the function $\psi(x) = 1$. Now we change the integration variables from $k_{\mu}$ to $q_{\mu}$:
\begin{align}
    k_{\mu}=\frac{q_{\mu}}{\sqrt{\tau}}-\frac{z_{\mu}}{2\tau},
\end{align}
which leads to:
\begin{align}\label{2hoperatorrep}
    h(\tau,x,y)=\frac{e^{-iz^{2}/(4\tau)}}{(4\pi\tau)^{d/2}}\int\frac{d^{d}q}{\pi^{d/2}}e^{iq^{2}}e^{-iH_{x}\tau+2\sqrt{\tau}q\cdot D-z\cdot D}1.
\end{align}
It is worth pointing out that in Eq.~\eqref{2hoperatorrep} the operators $H_{x}$ and $D_{\mu}$ should be considered to commute with $z_{\mu}$, i.e. the difference $z_{\mu} = x_{\mu}-y_{\mu}$ is held fixed in all subsequent calculations, and differential operators $\partial/\partial x^{\mu}$ do not act on $z_{\mu}$. 

At this point, we are not able to obtain the desired expansion in powers of $\tau$ by simply expanding the exponential of the operator because of the presence of the term $z^{\mu}D_{\mu}$ in Eq.~\eqref{2hoperatorrep}. It is possible to proceed by introducing the operator $L(\gamma)$:
\begin{align}
    L(\gamma) = e^{\gamma z\cdot D}e^{\gamma(-iH_{x}\tau+2\sqrt{\tau} q\cdot D-z\cdot D)}.
\end{align}
The operator $L(\gamma)$ satisfies the following differential equation:
\begin{align}\label{2Lequation}
    L^\prime(\gamma)=\left(-i\tau H_{x}(\gamma)+2\sqrt{\tau} q^{\mu}D_{\mu}(\gamma)\right)L(\gamma),
\end{align}
with the initial condition $L(0)=1$. In Eq.~\eqref{2Lequation} we defined new operators:
\begin{align}
    &H_{x}(\gamma) \equiv e^{\gamma z\cdot D}H_{x} e^{-\gamma z \cdot D} = D_{\mu}(\gamma)D^{\mu}(\gamma)+V(\gamma), \nonumber\\
    &D_{\mu}(\gamma)=e^{\gamma z\cdot D}D_{\mu}e^{-\gamma z \cdot D},\nonumber \\
    &V(\gamma) \equiv e^{\gamma z\cdot D}V(x)e^{-\gamma z \cdot D}.
\end{align}
We derive the explicit form of these operators as follows:
\begin{align}\label{2VDdefinition}
    &D_{\mu}(\gamma)=e^{\gamma z\cdot D}D_{\mu}e^{-\gamma z\cdot D}=D_{\mu}+\gamma z^{\alpha}[D_{\alpha},D_{\mu}]\nonumber\\ 
    &+\frac{\gamma^{2}}{2!}z^{\alpha}z^{\beta}[D_{\alpha},[D_{\beta},D_{\mu}]] +\cdots =D_{\mu}+i e\gamma z^{\alpha}F_{\alpha\mu}(x)\nonumber\\ 
    &+\frac{ie \gamma^{2}}{2!}z^{\alpha}z^{\beta}\partial_{\alpha}F_{\beta\mu}(x) +\cdots=D_{\mu}+ie\int_{0}^{\gamma}d\theta\big(z^{\alpha}F_{\alpha\mu}(x)\nonumber \\
    &+\theta z^{\beta}\partial_{\beta}z^{\alpha}F_{\alpha \mu}(x)+\cdots\big)=D_{\mu}+ ie z^{\alpha}\int_{0}^{\gamma}d\theta F_{\alpha \mu}(x+\theta z),\nonumber \\
    &V(\gamma) = e^{\gamma z\cdot D}V(x)e^{-\gamma z \cdot D} = V(x+\gamma z),
\end{align}
where $F_{\mu\nu}(x) = \partial_{\mu}A_{\nu}(x)-\partial_{\nu}A_{\mu}(x)=\frac{1}{ie}[D_{\mu},D_{\nu}]$.

The solution to the equation \eqref{2Lequation} reads:
\begin{align}
    L(\gamma) = \bm{P}\exp\left\{-i\tau\int_{0}^{\gamma}d\theta H_{x}(\theta)+2\sqrt{\tau}q^{\mu}\int_{0}^{\gamma}d\theta D_{\mu}(\theta)\right\},
\end{align}
where the symbol $\bm{P}$ denotes ordering of the operators with respect to the parameter $\theta$. Thus, we have the following representation for $h(\tau,x,y)$:
\begin{align}\label{2hbeforeq}
    &h(\tau,x,y)=\frac{e^{-iz^{2}/(4\tau)}}{(4\pi\tau)^{d/2}}e^{-z\cdot D}\int\frac{d^{d}q}{\pi^{d/2}}e^{iq^{2}}\nonumber\\
    &\times\bm{P}\exp\left\{-i\tau\int_{0}^{1}d\theta H_{x}(\theta)+2\sqrt{\tau}q^{\mu}\int_{0}^{1}d\theta D_{\mu}(\theta)\right\}1,
\end{align}
which allows us to integrate over $q_{\mu}$ with the help of the formula:
\begin{align}
    \int\frac{d^{d}q}{\pi^{d/2}}e^{iq^{2}+2q^{\mu} a_{\mu}}= i e^{-i\pi d/4}e^{ia^{2}},
\end{align}
and we obtain:
\begin{align}\label{2hPrepr}
    &h(\tau,x,y)=\frac{ie^{-iz^{2}/(4\tau)}}{(4\pi i\tau)^{d/2}}e^{-z\cdot D}\bm{P}\exp\Bigg\{-i\tau\bigg(\int_{0}^{1}d\theta H_{x}(\theta)\nonumber\\ 
    &- \int_{0}^{1}d\theta_{1}D_{\mu}(\theta_{1})\int_{0}^{1}d\theta_{2}D^{\mu}(\theta_{2})\bigg)\Bigg\}1.
\end{align}

A further simplification of the expression \eqref{2hPrepr} concerns disentangling the operator $e^{-z\cdot D}$. Namely, we define the operator $B(\gamma)$ as follows:
\begin{align}
    B(\gamma) = e^{-\gamma z\cdot D}e^{\gamma z \cdot \partial} = e^{-
    \gamma z\cdot \partial - i\gamma e z\cdot A(x)}e^{\gamma z\cdot \partial},
\end{align}
where $\partial_{\mu} \equiv \partial/\partial x^{\mu}$. Note that the operator $B(\gamma)$ satisfies the following equation:
\begin{align}
    &B^{\prime}(\gamma) = B(\gamma)e^{-\gamma z\cdot \partial}(-iez \cdot A(x))e^{\gamma z\cdot \partial}\nonumber\\ 
    &=-i e z^{\mu}B(\gamma)A_{\mu}(x-\gamma z),
\end{align}
which has the simple solution satisfying the initial condition $B(0) = 1$:
\begin{align}\label{2Bexpression}
    B(\gamma) = e^{-ie z^{\mu}\int_{0}^{\gamma}d\theta A_{\mu}(x-\theta z)}.
\end{align}
Using the expression \eqref{2Bexpression} we finally represent the heat kernel $h(\tau,x,y)$ as:
\begin{align}\label{2hfinalrepr}
    &h(\tau,x,y)=\frac{ie^{-iz^{2}/(4\tau)}}{(4\pi i\tau)^{d/2}}e^{-ie z^{\mu}\int_{0}^{1}d\theta A_{\mu}(x-\theta z)}e^{-z\cdot \partial}\nonumber\\ 
    &\times\bm{P}\exp\Bigg\{-i\tau\bigg(\int_{0}^{1}d\theta H_{x}(\theta)- \int_{0}^{1}d\theta_{1}D_{\mu}(\theta_{1})\nonumber\\ 
    &\times\int_{0}^{1}d\theta_{2}D^{\mu}(\theta_{2})\bigg)\Bigg\}1.
\end{align}
The action of the operator $e^{-z\cdot \partial}$ simply shifts the variable: $x_{\mu}\rightarrow x_{\mu}-z_{\mu}$. 

We see that in order to obtain the expansion in powers of $\tau$ from  Eq.~\eqref{2hfinalrepr} we need to expand the ordered exponential up to the desired order. Details about the usage of the ordered exponential of Eq.~\eqref{2hfinalrepr} can be found in the Appendix A.

Now we should present the result we obtained for the expansion of $h(\tau,x,y)$ up to the second order in $\tau$. After the computation of the action of the ordered exponential on $1$, the heat kernel $h(\tau,x,y)$ can be written in the following form:
\begin{widetext}
\begin{align}
    &h(\tau,x,y)=\frac{ie^{-iz^{2}/(4\tau)}}{(4\pi i\tau)^{d/2}}e^{-ie z^{\mu}\int_{0}^{1}d\theta A_{\mu}(y+\theta z)}\left[b_{0}(x,y)+b_{1}(x,y)\tau+b_{2}(x,y)\tau^{2}+O(\tau^{3})\right].
\end{align}
The functions $b_{n}(x,y)$ are gauge invariant. The zeroth-order coefficient is $b_{0}(x,y)=1$, and up to the second order we have:
\begin{align}\label{2b1}
    &b_{1}(x,y)= -i\int_{0}^{1}ds\,V(x_{s}) +e\int_{0}^{1}ds\,\partial_{\mu}B^{\mu}(s)+ie^{2}\int_{0}^{1}ds\,B_{\mu}(s)B^{\mu}(s) - ie^{2}\int_{0}^{1}ds_{1}B_{\mu}(s_{1})\int_{0}^{1}ds_{2}B^{\mu}(s_{2})\nonumber \\ 
    &=-i\int_{0}^{1}ds\, V(x_{s}) + ez^{\alpha}\int_{0}^{1}ds\, s(1-s) \partial_{\mu}F_{\alpha}^{\,\,\mu}(x_{s})-ie^{2}z^{\alpha}z^{\beta}\int_{0}^{1}ds_{1}\,s_{1}\int_{0}^{1}ds_{2}\,s_{2}F_{\alpha}^{\,\,\mu}(x_{s_{1}})F_{\beta\mu}(x_{s_{2}})\nonumber\\ 
    & + 2ie^{2}z^{\alpha}z^{\beta}\int_{0}^{1}ds_{1}\int_{0}^{s_{1}}ds_{2}\, s_{2}F_{\alpha}^{\,\,\mu}(x_{s_{1}})F_{\beta\mu}(x_{s_{2}}),
\end{align}
\begin{align}\label{2b2}
    &b_{2}(x,y)=-\frac{1}{2}\left[\int_{0}^{1}ds\,V(x_{s})+e^{2}\int_{0}^{1}ds_{1} B_{\mu}(s_{1})\int_{0}^{1}ds_{2}B^{\mu}(s_{2})-e^{2}\int_{0}^{1}ds \, B_{\mu}(s)B^{\mu}(s) -ie\int_{0}^{1}ds\,(1-2s) \partial_{\mu}B^{\mu}(s)\right]^{2}\nonumber\\ 
    &+ \frac{e^{2}}{12}F_{\mu \nu}(y)F^{\mu \nu}(y)+ 2ie\left(\int_{0}^{1}ds\, B^{\nu}(s)\right)\bigg[\int_{0}^{1}ds_{1}\, (1-s_{1})\partial_{\nu}V(x_{s_{1}})+2e^{2}\int_{0}^{1}ds_{1}(1-s_{1})\,\partial_{\nu}B^{\mu}(s_{1})\int_{0}^{1}ds_{2}B_{\mu}(s_{2})\nonumber\\ 
    &-2e^{2}\int_{0}^{1}ds_{1}(1-s_{1})B^{\mu}(s_{1})\partial_{\nu}B_{\mu}(s_{1})-ie\int_{0}^{1}ds_{1}(1-s_{1})(1-2s_{1})\partial_{\nu}\partial_{\mu}B^{\mu}(s_{1})\bigg] - \int_{0}^{1}ds\, s(1-s)\partial_{\nu}\partial^{\nu}V(x_{s})\nonumber\\ 
    &-2e^{2}\int_{0}^{1}ds_{1}s_{1}(1-s_{1})\partial_{\nu}\partial^{\nu}B^{\mu}(s_{1})\int_{0}^{1}ds_{2}B_{\mu}(s_{2})+ 2e^{2}\int_{0}^{1} ds \, s(1-s)\partial^{\nu}B^{\mu}(s)\partial_{\nu}B_{\mu}(s) \nonumber\\ 
    &+ 2e^{2}\int_{0}^{1}ds\, s(1-s)B^{\mu}(s)\partial_{\nu}\partial^{\nu}B_{\mu}(s)+ie\int_{0}^{1}ds\, s(1-s)(1-2s)\partial_{\nu}\partial^{\nu}\partial^{\mu}B_{\mu}(s)+2e^{2}F_{\mu\nu}(y)\int_{0}^{1}ds \, s(1-s)\partial^{\nu}B^{\mu}(s) \nonumber\\
    &+e^{2}\partial^{\nu}F_{\mu \nu}(y)\int_{0}^{1}ds\,\left(2s-2s^{2}-\frac{1}{3}\right)B^{\mu}(s)-4e^{2}\int_{0}^{1}ds_{2}\int_{0}^{s_{2}}ds_{1}s_{1}(1-s_{2})\partial^{\nu}B^{\mu}(s_{1})\partial_{\nu}B_{\mu}(s_{2})\nonumber \\ 
    &-4e^{2}\int_{0}^{1}ds_{2}\int_{0}^{s_{2}}ds_{1}s_{1}(1-s_{2})\partial^{\nu}B^{\mu}(s_{1})\partial_{\mu}B_{\nu}(s_{2})-e^{2}\int_{0}^{1}ds_{2}\int_{0}^{s_{2}}ds_{1}(1-2s_{1})B_{\nu}(s_{2})\partial^{\nu}\partial^{\mu}B_{\mu}(s_{1})\nonumber \\ 
    &-2ie\int_{0}^{1}ds_{2}\int_{0}^{s_{2}}ds_{1}B_{\mu}(s_{2})\partial^{\mu}V(x_{s_{1}})-2ie^{3}F_{\mu \nu}(y)\int_{0}^{1}ds_{2}\int_{0}^{s_{2}}ds_{1}B^{\mu}(s_{2})B^{\nu}(s_{1})\left(1+2s_{1}-2s_{2}\right) \nonumber\\ 
    &+ 4ie^{3}\int_{0}^{1}ds_{2}\int_{0}^{s_{2}}ds_{1}B_{\mu}(s_{1})\partial^{\nu}B^{\mu}(s_{1})B_{\nu}(s_{2})-4ie^{3}\int_{0}^{1}ds_{3}\int_{0}^{s_{3}}ds_{2}\int_{0}^{s_{2}}ds_{1}\bigg[\partial^{\nu}B_{\mu}(s_{1})B^{\mu}(s_{2})B_{\nu}(s_{3})\nonumber\\ 
    &+ \partial^{\nu}B_{\mu}(s_{1})B^{\mu}(s_{3})B_{\nu}(s_{2})+\partial^{\nu}B_{\mu}(s_{2})B^{\mu}(s_{1})B_{\nu}(s_{3})\bigg],
\end{align}
where we use notations $(x_{s})_{\mu}\equiv y_{\mu}+s z_{\mu}$, $B_{\mu}(s) \equiv z^{\alpha}\int_{0}^{s}d\theta F_{\alpha\mu}(x_{\theta})$, $\partial_{\mu}\equiv\partial/\partial y^{\mu}$. The differentiation with respect to $y_{\mu}$ should be done with $z_{\mu}$ held fixed. Our approach is not limited to the calculation of $b_{1}(x,y)$ and $b_{2}(x,y)$. Computation of the subsequent coefficients reduces to straightforward manipulations involving differential operators.
\end{widetext}

\section{Plane wave and constant fields}

In this section we verify that our results from Eqs.~\eqref{2b1} and~\eqref{2b2} reproduce correct small-time asymptotics of the heat kernels for two specific choices of the fields $A_{\mu}(x)$ and $V(x)$.
\subsection{Plane wave fields}
In this case the fields can be written as (see \cite{itzykson2006quantum}):
\begin{align}
    A_{\mu}(x)= \varepsilon_{\mu}f(\xi), \, V(x) = m^{2},    
\end{align}
where $\xi \equiv n^{\mu}x_{\mu}$, $\varepsilon_{\mu}$ is the polarization vector normalized as $\varepsilon_{\mu}\varepsilon^{\mu}=-1$, $n_{\mu}$ is a lightlike vector ($n_{\mu}n^{\mu}=0$) that specifies the direction of wave propagation. The vectors $\varepsilon_{\mu}$ and $n_{\mu}$ are orthogonal: $\varepsilon_{\mu}n^{\mu}=0$. The field strength tensor is $F_{\mu \nu}(x) = \phi_{\mu \nu}f^{\prime}(\xi)$, where $\phi_{\mu \nu}\equiv n_{\mu}\varepsilon_{\nu}-n_{\nu}\varepsilon_{\mu}$. Hereafter the function $f(\xi)$ is arbitrary (not only $e^{i\xi}$).

The heat kernel $h(\tau,x,y)$ for the plane wave fields is well-known. A derivation of the closed-form expression suitable for QED can be found, for example, in \cite{itzykson2006quantum, schwinger1951gauge}. One can similarly obtain the expression for scalar electrodynamics:
\begin{align}\label{3hplane}
    &h(\tau,x,y) = -\frac{i}{(4\pi \tau)^{2}}\exp \Big\{-i\Big[\frac{z^{2}}{4\tau}+m^{2}\tau+e^{2}\langle\delta f^{2}\rangle\tau\nonumber\\ 
    &+ez^{\mu}\int_{0}^{1}d\theta \,A_{\mu}(y+\theta z)\Big]\Big\},
\end{align}
where
\begin{align}
    \langle\delta f^{2}\rangle = \int_{\xi^{\prime}}^{\xi}\frac{ds\,f^{2}(s)}{\xi - \xi'}-\left(\int_{\xi'}^{\xi}\frac{ds \, f(s)}{\xi-\xi'}\right)^{2},
\end{align}
with $\xi \equiv n^{\mu}x_{\mu}$, $\xi'\equiv n^{\mu}y_{\mu}$. From Eq.~\eqref{3hplane} we get the small-time expansion:
\begin{align}\label{3hsmalltime}
    &h(\tau,x,y) = -\frac{ie^{-iz^{2}/4\tau}}{(4\pi \tau)^{2}}e^{-iez^{\mu}\int_{0}^{1}d\theta \,A_{\mu}\left(y+\theta z\right)}\\ \nonumber
    &\times\Big[1-i\tau\left(m^{2}+e^{2}\langle\delta f^{2}\rangle\right)-\frac{\tau^{2}}{2}\left(m^{2}+e^{2}\langle\delta f^{2}\rangle\right)^{2}+O(\tau^{3})\Big].
\end{align}

Obtaining the same asymptotics with our approach involves using cumbersome expression for $b_{2}(x,y)$. Fortunately, most of the terms in Eq.~\eqref{2b2} vanish upon substituting the vector $B_{\mu}(s)$ for the plane wave fields case. This vector reads:
\begin{align}\label{3Bexpr}
    &B_{\mu}(s) = z^{\alpha}\int_{0}^{s}d\theta \, F_{\alpha \mu}(y+\theta z) = z^{\alpha}\phi_{\alpha\mu}\nonumber\\ 
    &\times\int_{0}^{s}d\theta \, f^{\prime}(\xi^{\prime}+\theta(\xi-\xi^{\prime})) = \frac{z^{\alpha}\phi_{\alpha \mu}}{\xi - \xi^{\prime}}[f(\xi^{\prime}+s(\xi-\xi^{\prime}))\\ \nonumber
    &-f(\xi^{\prime})].
\end{align}
Therefore, differentiation of $B_{\mu}(s)$ with respect to $y^{\nu}$ always produces the vector $n^{\nu}$:
\begin{align}\label{3dytodxi}
    \partial_{\nu}B_{\mu}(s) = n_{\nu}\partial_{\xi^{\prime}}B_{\mu}(s),
\end{align}
where $\partial_{\xi^{\prime}}\equiv \partial/\partial\xi^{\prime}$. It is also worth recalling that in Eqs.~\eqref{2b1} and~\eqref{2b2} differentiation with respect to $y^{\nu}$ should be done with $z^{\mu}$ held fixed. Likewise, the differentiation with respect to $\xi^{\prime}$ in Eq.~\eqref{3dytodxi} should be done with $\xi - \xi^{\prime}$ fixed. The appearance of $n^{\nu}$ in Eq.~\eqref{3dytodxi} and identities $\phi_{\mu \nu}\phi_{\alpha}^{\,\,\,\mu}\phi_{\beta}^{\,\,\,\nu} = 0$, $n^{\mu}\phi_{\mu\nu}=0$ greatly simplify the expression for $b_{2}(x,y)$. A careful inspection shows that the only non-vanishing terms are as follows:
\begin{align}
    &b_{1}(x,y) = -i\int_{0}^{1}ds\,V(x_{s})+ie^{2}\int_{0}^{1}ds\,B_{\mu}(s)B^{\mu}(s)\nonumber\\ 
    &-ie^{2}\int_{0}^{1}ds_{1}B_{\mu}(s_{1})\int_{0}^{1}ds_{2}B^{\mu}(s_{2}),
\end{align}
\begin{align}
    &b_{2}(x,y)=-\frac{1}{2}\bigg[\int_{0}^{1}ds\,V(x_{s})-e^{2}\int_{0}^{1}ds\,B_{\mu}(s)B^{\mu}(s)\nonumber\\ 
    &+e^{2}\int_{0}^{1}ds_{1}B_{\mu}(s_{1})\int_{0}^{1}ds_{2}B^{\mu}(s_{2})\bigg]^{2}.
\end{align}
Using Eq.~\eqref{3Bexpr}, it is then straightforward to show that:
\begin{align}
    &\int_{0}^{1}ds_{1}B_{\mu}(s_{1})\int_{0}^{1}ds_{2}B^{\mu}(s_{2}) - \int_{0}^{1}ds\,B_{\mu}(s)B^{\mu}(s)\nonumber\\ 
    &= \langle\delta f^{2}\rangle,
\end{align}
so upon substituting $V(x_{s}) = m^{2}$ we get a result that reproduces the corresponding terms in Eq.~\eqref{3hsmalltime}:
\begin{align}
    &b_{1}(x,y) = -i\left(m^{2}+e^{2}\langle\delta f^{2}\rangle\right),\nonumber\\ 
    &b_{2}(x,y) = -\frac{1}{2}\left(m^{2}+e^{2}\langle\delta f^{2}\rangle\right)^{2}.
\end{align}

\subsection{Constant fields}

The heat kernel for constant fields $F_{\mu\nu}(x) = F_{\mu\nu}$, $V(x)=m^{2}$ can be written using matrix notation as (see \cite{itzykson2006quantum}):
\begin{align}\label{3hconstant}
    &h(\tau,x,y) = -\frac{i}{(4\pi \tau)^{2}}e^{-\mathcal{L}(\tau)}\exp\bigg\{-iez^{\mu}\int_{0}^{1}d\theta\nonumber\\ 
    &\times  \, A_{\mu}(y+\theta z)- \frac{i}{4}z\left[eF\,\mathrm{cth}\left(eF\tau\right)\right]z-im^{2}\tau\bigg\},
\end{align}
where:
\begin{align}
\mathcal{L}(\tau)= \frac{1}{2}\mathrm{Tr}\, \ln \left[(eF\tau)^{-1}\mathrm{sh}\left(eF\tau\right)\right]. 
\end{align}
The matrix expressions should be understood, for example, as:
\begin{align}
z\left[eF\,\mathrm{cth}\left(eF\tau\right)\right]z = ez^{\alpha}F_{\alpha}^{\,\,\beta}\left(\mathrm{cth\left(eF\tau\right)}\right)_{\beta\gamma}z^{\gamma}.
\end{align}
The functions of the matrix $F$ are defined by the Taylor series, and powers of $F$ should be understood as $(F^{n})_{\alpha}^{\,\,\,\beta} = F_{\alpha}^{\,\,\,\beta_{1}}F_{\beta_{1}}^{\,\,\,\beta_{2}}...F_{\beta_{n-1}}^{\,\,\,\beta}$. For the expansion of the heat kernel \eqref{3hconstant} as $\tau \to0$ we use the following formulae:
\begin{align}
    &\mathcal{L}(\tau) = \frac{1}{2}\mathrm{Tr}\, \ln \left[(eF\tau)^{-1}\mathrm{sh}\left(eF\tau\right)\right] = \frac{1}{2}\mathrm{Tr}\, \ln \bigg[(eF\tau)^{-1}\nonumber\\ 
    &\times\left(eF\tau+\frac{1}{6}(eF\tau)^{3}+O(\tau^{5})\right)\bigg] = \frac{e^{2}\tau^{2}}{12}\mathrm{Tr}\,F^{2} + O(\tau^{4}),\nonumber\\ 
    &eF\,\mathrm{cth}\left(eF\tau\right) = \frac{1}{\tau}+\frac{1}{3}(eF)^{2}\tau+O(\tau^{3}).
\end{align}
The small-time expansion of Eq.~\eqref{3hconstant} up to the second order then reads:
\begin{align}\label{3hconstexp}
    &h(\tau,x,y) = -\frac{ie^{-iz^{2}/(4\tau)}}{(4\pi\tau)^{2}}e^{-ie z^{\mu}\int_{0}^{1}d\theta A_{\mu}(y+\theta z)}\nonumber\\ 
    &\times\bigg[1-i\tau\left(m^{2}+\frac{e^{2}}{12}zF^{2}z\right)-\frac{e^{2}\tau^{2}}{12}\mathrm{Tr}\,F^{2}\nonumber\\ 
    &-\frac{\tau^{2}}{2}\left(m^{2}+\frac{e^{2}}{12}zF^{2}z\right)^{2}+O(\tau^{3})\bigg].
\end{align}

Let us obtain this result from Eqs.~\eqref{2b1} and~\eqref{2b2}. The vector $B_{\mu}(s) = z^{\alpha}\int_{0}^{s}d\theta\,F_{\alpha\mu} = sz^{\alpha}F_{\alpha\mu}$ does not depend on $x_{\mu}$ and $y_{\mu}$. Therefore, most of the terms in Eqs.~\eqref{2b1} and~\eqref{2b2} vanish again. Substituting $B_{\mu}(s) = sz^{\alpha}F_{\alpha\mu}$ we get in the first order:
\begin{align}
    &b_{1}(x,y) = -im^{2} - ie^{2}z^{\alpha}z^{\beta}F_{\alpha\mu}F_{\beta}^{\,\,\mu}\int_{0}^{1}ds_{1}s_{1}\int_{0}^{1}ds_{2}s_{2} \nonumber\\
    &+ie^{2}z^{\alpha}z^{\beta}F_{\alpha\mu}F_{\beta}^{\,\,\mu}\int_{0}^{1}ds\,s^{2} = -im^{2}+i\frac{e^{2}}{12}z^{\alpha}z^{\beta}F_{\alpha\mu}F_{\beta}^{\,\,\mu} \nonumber\\ 
    &= -im^{2}-i\frac{e^{2}}{12}zF^{2}z,
\end{align}
which coincides with the corresponding coefficient in Eq.~\eqref{3hconstexp}. Similarly, for the second order we obtain from Eq.~\eqref{2b2}:
\begin{align}
    &b_{2}(x,y) = -\frac{1}{2}\left(m^{2}-\frac{e^{2}}{12}z^{\alpha}z^{\beta}F_{\alpha\mu}F_{\beta}^{\,\,\mu}\right)^{2}+\frac{e^{2}}{12}F_{\mu\nu}F^{\mu\nu} \nonumber\\
    &=-\frac{1}{2}\left(m^{2}+\frac{e^{2}}{12}zF^{2}z\right)^{2}-\frac{e^{2}}{12}\mathrm{Tr}\,F^{2},
\end{align}
which also reproduces the corresponding coefficient from Eq.~\eqref{3hconstexp}.

\section{Conclusion}
We have described a new method for computing small-time expansion of the heat kernel $h(\tau,x,y)$ for the Klein-Gordon operator in flat space-time coupled to electromagnetic fields. This method can be considered as an extension of the approach described in \cite{nepomechie1985calculating}. Our extension allows one to compute expansion coefficients for arbitrary space-time points $x_{\mu}$ and $y_{\mu}$. It is based on the representation \eqref{2hfinalrepr}, which reduces the problem to straightforward manipulations involving differential operators acting on the function $\psi(x)=1$. We obtained closed-form expressions for the expansion coefficients up to the second order in $\tau$. Our results correctly reproduce the small-time asymptotics of the heat kernel for two simple configurations: a plane wave field and constant fields.

\begin{acknowledgments}
The authors are grateful to R.N. Lee for helpful discussions and comments.
\end{acknowledgments}

\appendix

\section{Details of the ordered exponential computation}

In Section 2 we obtain a representation of the heat kernel (see Eq.~\eqref{2hfinalrepr}) that contains the following ordered exponential:
\begin{align}\label{APorder}
    &\bm{P}\exp\bigg\{-i\tau\bigg(\int_{0}^{1}d\theta H_{x}(\theta)-\int_{0}^{1}d\theta_{1}D_{\mu}(\theta_{1})\nonumber\\
    &\times\int_{0}^{1}d\theta_{2}D^{\mu}(\theta_{2})\bigg)\bigg\},
\end{align}
where $H_{x}(\theta) = D_{\mu}(\theta)D^{\mu}(\theta)+V(\theta)$, and the operators $D_{\mu}(\theta)$ and $V(\theta)$ are defined in Eqs.~\eqref{2VDdefinition}. Because of the presence of the double integral, Eq.~\eqref{APorder} differs from the usual ordered exponential, which contains only a single integral. Therefore, it is not obvious whether the conventional rules for working with the ordering operation apply to Eq.~\eqref{APorder}. In this section we demonstrate an interpretation of the ordering operation that leads to the correct results.

% It is worth to note from the beginning that the double integral from Eq.~\eqref{APorder} can be written in another form:
% \begin{align}\label{Adoubleint}
%     &\int_{0}^{1}d\theta_{1}D_{\mu}(\theta_{1})\int_{0}^{1}d\theta_{2}D^{\mu}(\theta_{2})\nonumber\\ 
%     &= 2\int_{0}^{1}d\theta_{1}D_{\mu}(\theta_{1})\int_{0}^{\theta_{1}}d\theta_{2}D^{\mu}(\theta_{2}).
% \end{align}
% Both forms in Eq.~\eqref{Adoubleint} give the same result if one follows the rules stated below. In the following we only use the first form: $\int_{0}^{1}d\theta_{1}D_{\mu}(\theta_{1})\int_{0}^{1}d\theta_{2}D^{\mu}(\theta_{2})$.

Let us demonstrate the use of the operator \eqref{APorder} by expanding it up to the second order in $\tau$, i.e. up to the term proportional to $\tau^{2}$. Expressing the exponential as the power series we get:
\begin{widetext}
    
\begin{align}\label{APexpansion}
    &\bm{P}\exp\Bigg\{-i\tau\bigg(\int_{0}^{1}d\theta H_{x}(\theta)- \int_{0}^{1}d\theta_{1}D_{\mu}(\theta_{1})\int_{0}^{1}d\theta_{2}D^{\mu}(\theta_{2})\bigg)\Bigg\} = 1 -i\tau\bm{P}\Bigg\{\int_{0}^{1}d\theta H_{x}(\theta) -\int_{0}^{1}d\theta_{1}D_{\mu}(\theta_{1})\nonumber\\ 
    &\times\int_{0}^{1}d\theta_{2}D^{\mu}(\theta_{2})\Bigg\} -\frac{\tau^{2}}{2}\bm{P}\Bigg \{\int_{0}^{1}d\theta_{1}H_{x}(\theta_{1})\int_{0}^{1}d\theta_{2}H_{x}(\theta_{2})-2\int_{0}^{1}d\theta_{1}H_{x}(\theta_{1})\int_{0}^{1}d\theta_{2}D_{\mu}(\theta_{2})\int_{0}^{1}d\theta_{3}D^{\mu}(\theta_{3})\nonumber\\ 
    &+\int_{0}^{1}d\theta_{1}D_{\mu}(\theta_{1})\int_{0}^{1}d\theta_{2}D^{\mu}(\theta_{2})\int_{0}^{1}d\theta_{3}D_{\nu}(\theta_{3})\int_{0}^{1}d\theta_{4}D^{\nu}(\theta_{4})\Bigg\}+O(\tau^{3}).
\end{align}
First, we claim that when working with sums of operators as in Eq.~\eqref{APexpansion}, the ordering operation $\bm{P}$ should be considered linear. Second, the following standard definition should be used:
\begin{align}
    &\bm{P}\Big\{A_{1}(\gamma_{1})\cdots A_{n}(\gamma_{n})\Big\} =\sum_{\pi \in S_{n}}\left(\prod_{j=1}^{n-1}\theta(\gamma_{\pi_{j}}-\gamma_{\pi_{j+1}})\right) A_{\pi_{1}}(\gamma_{\pi_{1}})A_{\pi_{2}}(\gamma_{\pi_{2}})\cdots A_{\pi_{n}}(\gamma_{\pi_{n}}),
\end{align}
where $\theta(x)$ is the Heaviside theta function and $S_{n}$ is the symmetric group of degree $n$. With these two rules we obtain:
\begin{align}\label{APexp}
    &\bm{P}\exp\Bigg\{-i\tau\bigg(\int_{0}^{1}d\theta H_{x}(\theta)- \int_{0}^{1}d\theta_{1}D_{\mu}(\theta_{1})\int_{0}^{1}d\theta_{2}D^{\mu}(\theta_{2})\bigg)\Bigg\} = 1 - i \tau \bigg(\int_{0}^{1}d\theta H_{x}(\theta)-2\int_{0}^{1}d\theta_{1}D_{\mu}(\theta_{1})\nonumber\\
    &\times\int_{0}^{\theta_{1}}d\theta_{2}D^{\mu}(\theta_{2})\bigg)-\tau^{2}\bigg(\int_{0}^{1}d\theta_{1}H_{x}(\theta_{1})\int_{0}^{\theta_{1}}d\theta_{2}H_{x}(\theta_{2})-2\int_{0}^{1}d\theta_{1}H_{x}(\theta_{1})\int_{0}^{\theta_{1}}d\theta_{2}D_{\mu}(\theta_{2})\int_{0}^{\theta_{2}}d\theta_{3}D^{\mu}(\theta_{3})\nonumber\\ 
    &-2\int_{0}^{1}d\theta_{1}D_{\mu}(\theta_{1})\int_{0}^{\theta_{1}}d\theta_{2}H_{x}(\theta_{2})\int_{0}^{\theta_{2}}d\theta_{3}D^{\mu}(\theta_{3}) - 2\int_{0}^{1}d\theta_{1}D_{\mu}(\theta_{1})\int_{0}^{\theta_{1}}D^{\mu}(\theta_{2})\int_{0}^{\theta_{2}}d\theta_{3}H_{x}(\theta_{3})+4\int_{0}^{1}d\theta_{1}D_{\mu}(\theta_{1})\nonumber\\
    &\times\int_{0}^{\theta_{1}}d\theta_{2}D^{\mu}(\theta_{2})\int_{0}^{\theta_{2}}d\theta_{3}D_{\nu}(\theta_{3})\int_{0}^{\theta_{3}}d\theta_{4}D^{\nu}(\theta_{4})+4\int_{0}^{1}d\theta_{1}D_{\mu}(\theta_{1})\int_{0}^{\theta_{1}}d\theta_{2}D_{\nu}(\theta_{2})\int_{0}^{\theta_{2}}d\theta_{3}D^{\mu}(\theta_{3})\int_{0}^{\theta_{3}}d\theta_{4}D^{\nu}(\theta_{4})\nonumber\\ 
    &+4\int_{0}^{1}d\theta_{1}D_{\mu}(\theta_{1})\int_{0}^{\theta_{1}}d\theta_{2}D_{\nu}(\theta_{2})\int_{0}^{\theta_{2}}d\theta_{3}D^{\nu}(\theta_{3})\int_{0}^{\theta_{3}}d\theta_{4}D^{\mu}(\theta_{4})\bigg)+O(\tau^{3}).
\end{align}

We verify that the interpretation of the ordered exponential described above is correct by recalling the representation \eqref{2hbeforeq}. The ordered exponential in Eq.~\eqref{2hbeforeq} does not contain double integrals and, therefore, can be interpreted in the standard way. Expanding the exponential from Eq.~\eqref{2hbeforeq} and collecting the terms with the same power of $\tau$:
\begin{align}
    &\bm{P} \exp\left\{-i\tau \int_{0}^{1}d\theta H_{x}(\theta) + 2\sqrt{\tau}q^{\mu}\int_{0}^{1}d\theta D_{\mu}(\theta)\right\}  = F_{0}+F_{1}\tau+F_{2}\tau^{2}+O(\tau^{3}),
\end{align}
we can then integrate the operators $F_{n}$ over $q_{\mu}$ and compare the results with the expansion \eqref{APexp}. Let us write out the expression for the $F_{2}$ term explicitly:
\begin{align}
    &F_{2} = -\int_{0}^{1}d\theta_{1}H_{x}(\theta_{1})\int_{0}^{\theta_{1}}d\theta_{2}H_{x}(\theta_{2})-4iq^{\alpha}q^{\beta}\bigg[\int_{0}^{1}d\theta_{1}H_{x}(\theta_{1})\int_{0}^{\theta_{1}}d\theta_{2}D_{\alpha}(\theta_{2})\int_{0}^{\theta_{2}}d\theta_{3}D_{\beta}(\theta_{3})+\int_{0}^{1}d\theta_{1}D_{\alpha}(\theta_{1})\nonumber\\
    &\times\int_{0}^{\theta_{1}}d\theta_{2}H_{x}(\theta_{2})\int_{0}^{\theta_{2}}d\theta_{3}D_{\beta}(\theta_{3})+\int_{0}^{1}d\theta_{1}D_{\alpha}(\theta_{1})\int_{0}^{\theta_{1}}d\theta_{2}D_{\beta}(\theta_{2})\int_{0}^{\theta_{2}}d\theta_{3}H_{x}(\theta_{3})\bigg]+16q^{\alpha}q^{\beta}q^{\mu}q^{\nu}\int_{0}^{1}d\theta_{1}D_{\alpha}(\theta_{1})\nonumber\\ 
    &\times\int_{0}^{\theta_{1}}d\theta_{2}D_{\beta}(\theta_{2})\int_{0}^{\theta_{2}}d\theta_{3}D_{\mu}(\theta_{3})\int_{0}^{\theta_{3}}d\theta_{4}D_{\nu}(\theta_{4}). 
\end{align}
After using the following formulae to perform integration over $q_{\mu}$:
\begin{align}
    &\int\frac{d^{d}q}{\pi^{d/2}}e^{iq^{2}}q_{\alpha}q_{\beta} = -\frac{1}{2}e^{-i\pi d/4}g_{\alpha \beta},\nonumber\\ 
    &\int\frac{d^{d}q}{\pi^{d/2}}e^{iq^{2}}q_{\alpha}q_{\beta}q_{\mu}q_{\nu} = -\frac{i}{4}e^{-i\pi d/4}(g_{\alpha \beta}g_{\mu \nu}+ g_{\alpha \mu}g_{\beta \nu}+g_{\alpha \nu}g_{\beta \mu}),
\end{align}
we see that integration of $F_{2}$ gives the same result as in Eq.~\eqref{APexp}, which confirms the rules we formulated above for interpreting the ordered exponential from Eq.~\eqref{APorder}.

\end{widetext}

\bibliography{apssamp}% Produces the bibliography via BibTeX.
\bibliographystyle{apsrev4-2}

\end{document}